\documentclass[manuscript]{acmart}

\usepackage{xcolor}
\usepackage{placeins}
\usepackage[percent]{overpic}

\AtBeginDocument{%
  }

\setcopyright{acmlicensed}
\copyrightyear{2018}
\acmYear{2027}
\usepackage[normalem]{ulem}
\newcommand{\etap}{\eta^2_p}

\begin{document}


\title{Evaluative Dynamics of AI Integration and Expert Performance under Epistemic Dependence across Heterogeneous Stakes}


\author{Dennis Kim}
\affiliation{%
  \institution{Colorado State University}
  \city{Colorado}
  \country{USA}}
\email{d.kim@colostate.edu}

\author{Roya Daneshi}
\affiliation{%
  \institution{Colorado State University}
  \city{Colorado}
  \country{USA}}
\email{roya.daneshi@colostate.edu}

\author{Nikhil Krishnaswamy}
\affiliation{%
  \institution{Colorado State University}
  \city{Colorado}
  \country{USA}}
\email{nkrishna@colostate.edu}

\author{Bruce Draper}
\affiliation{%
  \institution{Colorado State University}
  \city{Colorado}
  \country{USA}}
\email{Bruce.Draper@colostate.edu}

\author{Sarath Sreedharan}
\affiliation{%
  \institution{Colorado State University}
  \city{Colorado}
  \country{USA}}
\email{sarath.sreedharan@colostate.edu}


\begin{abstract}
AI is increasingly integrated into expert workflows, yet how integration affects perceptions of the expert, AI, and their combination remains unclear in domains where lay users are epistemically dependent on AI-assisted experts. We examine this through a novel controlled medical study ($N = 166$) and a direct cross-domain analysis with pre-existing academic-advising data ($n = 157$, combined $N = 323$). Expert errors reduced evaluations of the human expert across domains. Perceived expertise, however, varied by AI integration strategy in the higher-stakes medical task, where automatic AI oversight produced higher ratings than expert-only or expert-initiated AI. Exploratory ordinal sensitivity analyses identified a performance-contingent reuse pattern, with automatic oversight producing greater intended reuse after successful medical performance. Overall, performance-related recalibration appeared comparatively portable, while integration-structure effects were more selective and context-sensitive. These findings suggest that system designers should consider how AI enters expert workflows, not only whether it is present.
\end{abstract}

\maketitle

\section{Introduction}

AI is increasingly being incorporated into expert decision-making and advisory workflows, where users may rely solely on an expert, augmenting rather than replacing human experts \cite{panigutti2022understanding, mcnamara2024clinician}. In many such systems, the human expert still gives the recommendation while the AI provides monitoring or corrective support, such as checking or verification \cite{sridharan2026impact, topff2024artificial, dratsch2023automation}. These systems comprise multiple distinct actors: the AI, the human, and the human-AI team \cite{riedl2024patients, oneill2022hat}, which implies that evaluations may need to target each entity separately \cite{riedl2024patients, kim2026implications, bansal2019beyond, bansal2021does}. Trust itself is multi-dimensional rather than a single one-dimensional metric \cite{hendriks2015meti, oneill2022hat, kim2026implications}. How the AI enters the workflow may also matter: an expert who deliberately invokes AI may be interpreted differently from an AI system that automatically monitors and intervenes \cite{green2019principles}. Together, these features raise questions about how trust is distributed across the human, AI, and human-AI team as performance and AI integration structure vary.

Trust should calibrate to evidence about reliability \cite{lee2004trust, zhang2020confidence, bucinca2021trust}, and observed expert errors are among the strongest such evidence \cite{rechkemmer2022confidence, HanKo2025TrustDynamicsFA, dietvorst2015aversion}. AI-supported workflows complicate this calibration. When AI detects or corrects an expert's mistake, an expert error no longer necessarily implies a failed system outcome \cite{kim2026implications}, potentially separating evaluations of the expert from those of the overall system and making it unclear who bears the evaluative consequences of the original error \cite{lima2021moral}. Trust may remain concentrated on the expert, shift toward the AI, spill over to the human-AI team, or be redistributed across all three \cite{riedl2024patients, kim2026implications}.

The only direct test of this expert-AI workflow paradigm to date comes from academic advising \cite{kim2026implications}. There, advisor failure reduced advisor-directed trust even when AI corrected the error, AI-directed trust was comparatively insensitive to expert performance, and team-directed effects were inconsistent. These results are informative, but they were obtained in a setting where the cost of an uncorrected expert error is comparatively low. That matters because stakes are not a neutral backdrop for trust.

Not all contexts characterized by epistemic dependence are interchangeable. Although they all involve lay users relying on expertise they cannot independently verify, they differ in perceived stakes and professional expectations, factors that can shape trust in AI-assisted decision making \cite{castelo2019taskdependent, riedl2024patients}. These contextual differences may also change how users interpret expert errors and the meaning of visible assistance or help-seeking \cite{brooks2015advice, castelo2019taskdependent}.

Medicine is particularly important because patients often depend on specialized expertise they cannot independently verify, a form of high-stakes epistemic dependence \cite{hall2001trust, hardwig1985epistemic, hendriks2015meti}.  Trust in physicians accordingly involves expectations of competence, honesty, confidentiality, and fidelity to patients' interests \cite{hall2001trust}, while medical decisions carry strong expectations of careful judgment and professional responsibility \cite{asan2020aiTrustHealthcare, hall2001trust}. As AI is increasingly used for clinical decision support, diagnosis, reasoning, and other forms of assistance \cite{asan2020aiTrustHealthcare, moell2025medical, panigutti2022understanding}, its involvement may also carry social or competence signals. Prior work shows that visible AI involvement can be interpreted either as appropriate assistance or as evidence of over-reliance, reduced human effort or agency, or diminished competence \cite{nakano2025understanding}. Expert-initiated and automatic oversight may therefore carry different social meanings in medicine. Medicine therefore provides a more consequential context for testing how observable expert performance and AI integration structure shape user evaluations under epistemic dependence.


\begin{figure*}[!t]
    \centering
    \includegraphics[width=0.78\textwidth]{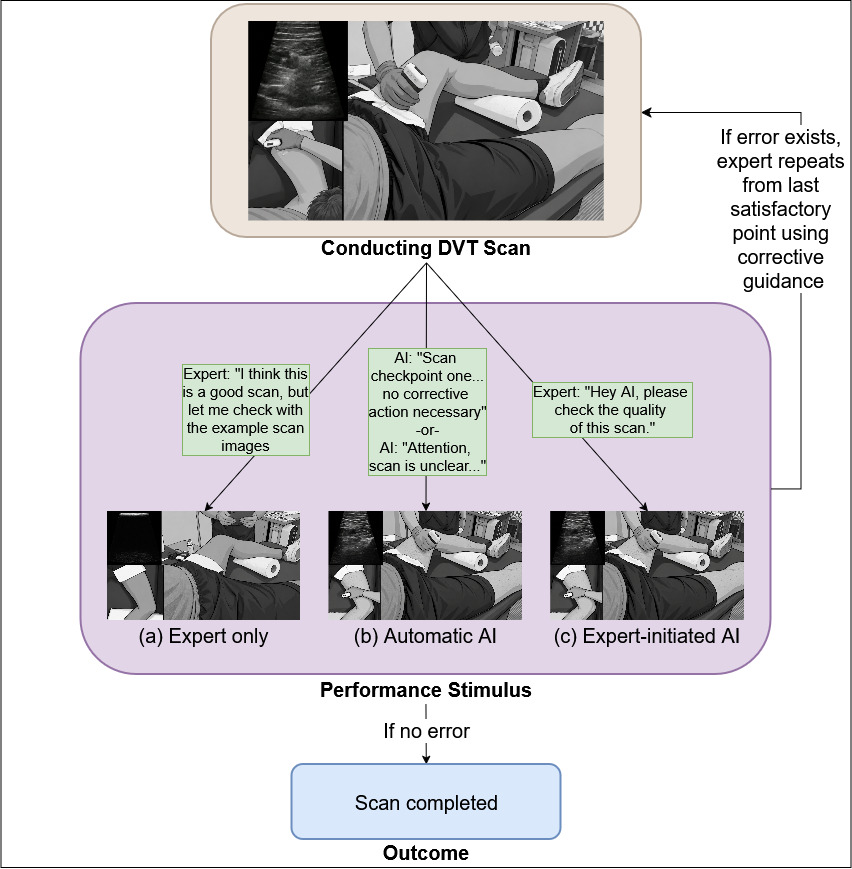}

    \caption{Medical study procedure and experimental conditions.
    Participants observed an expert conducting a DVT scan under one of three AI integration conditions:
    (a) Expert Only, in which the expert consulted reference material;
    (b) Automatic AI, in which AI continuously monitored the scan and provided corrective guidance when necessary; and
    (c) Expert-Initiated AI, in which the expert explicitly requested AI assistance.
    If an error occurred, the expert repeated the procedure from the last satisfactory point using the corresponding corrective guidance.
    In the no-error condition, the scan was completed without correction.}

    \Description{Diagram of the medical study procedure under Expert Only,
    Automatic AI, and Expert-Initiated AI conditions.}

    \label{fig:medical_workflows}
\end{figure*}

We address these issues through a new controlled user study of a simulated DVT (Deep Vein Thrombosis; refer to \cite{needleman2018ultrasound} for procedural details) ultrasound workflow in which participants evaluated a technician under expert-only, automatic-AI, or expert-initiated-AI conditions. Figure~\ref{fig:medical_workflows} illustrates the three workflow structures we consider, which differ in how the expert's work is checked and corrected (in the presence of an error) as well as their performance. We manipulated expert performance and examined multidimensional evaluations and intended reuse of the technician and, where applicable, trust in the AI and human-AI team. We then combined the newly collected medical data with the pre-existing advising data \cite{kim2026implications} to conduct a direct cross-domain analysis of expert performance and AI integration structure. Rather than treating the advising findings as generally applicable, this analysis tests which evaluative patterns persist across substantially different contexts of epistemic dependence and which instead reveal domain-specific boundary conditions.

Observable expert errors provide direct evidence about the expert's performance and should therefore reduce evaluations of the expert even when the overall workflow ultimately succeeds. \textbf{H1: Expert errors will reduce evaluations of the human expert, including perceived expertise, integrity, benevolence, expert-directed trust, general trust, and intended reuse.} When AI detects or corrects an expert error, the observable failure nevertheless originates with the human rather than the AI. The evaluative consequences of that failure may therefore remain concentrated on the expert rather than transfer to the corrective AI. \textbf{H2: Expert errors will reduce trust in the expert, while trust in the corrective AI assistant will be comparatively less sensitive to expert performance.} The implications for the combined human-AI team are less clear because the team includes both the human who makes the error and the AI that detects it and provides corrective guidance. \textbf{RQ1: To what extent do expert errors also affect trust in the human-AI team?} The advising study \cite{kim2026implications} cannot establish whether its performance-related effects persist beyond that comparatively lower-stakes setting or instead depend on the domain in which expert reliance occurs. Because reactions to, and reliance on, algorithms are task-dependent \cite{castelo2019taskdependent, salimzadeh2024dealing}, we directly test this boundary across advising and medicine. \textbf{RQ2: Does the effect of expert performance on trust-related evaluations differ between academic advising and medicine?} Additionally, the meaning of how AI enters the interaction may itself be domain-dependent. Seeking assistance can shape perceptions of the person seeking it, with advice seeking sometimes increasing perceived competence while reliance on computerized decision support can produce less favorable evaluations \cite{brooks2015advice, shaffer2013derogate}. If these interpretations vary by domain, the effect of AI modality could vary as well. \textbf{RQ3: Does the effect of AI intervention modality on trust-related evaluations differ between academic advising and medicine?} Finally, the effect of expert performance may itself depend on how AI enters the workflow. Because automatic and expert-initiated support differ in how assistance is initiated and presented, the same expert error may carry different procedural meaning across domains. \textbf{RQ4: Does domain alter whether AI intervention modality moderates the effect of expert performance on trust-related evaluations?}

This paper makes the following contributions:

\begin{itemize}
    \item We provide, to the best of our knowledge, the first controlled evaluation of this expert-AI workflow paradigm in a more consequential medical context, establishing how expert performance and AI integration structure shape user evaluations in that setting.
    
    \item We move beyond the prior single-domain evidence by combining the newly collected medical data with the pre-existing academic-advising data from Kim et al.~\cite{kim2026implications} and directly modeling Domain across the two settings. This cross-domain analysis distinguishes comparatively portable performance-related penalties from more selective, domain-sensitive effects of AI integration structure, thereby identifying boundary conditions that cannot be established from single-domain evidence alone.
\end{itemize}

The medical study first showed that AI Modality significantly affected perceived expertise, general trust, intended reuse, affective comfort toward the technician, and trust in technician-provided information. The cross-domain analyses then showed that expert errors reduced all eight shared human-facing outcomes, supporting H1, while human-directed trust was performance-sensitive and AI-directed trust was comparatively insensitive in AI-supported conditions, consistent with H2. Team-directed effects were nonsignificant (RQ1), and performance effects did not differ systematically by domain (RQ2). Perceived expertise showed a significant Domain $\times$ Modality interaction (RQ3). No Domain $\times$ Modality $\times$ Performance interaction reached significance in the primary ANOVAs (RQ4), although exploratory analyses identified an automatic-AI reuse advantage after successful medical performance.

\section{Related Work}

\subsection{Generalizing Human-AI Trust Across Expert Domains}

Trust in AI-supported decision making is shaped not only by the characteristics of an AI system, but also by the context in which that system is used. Foundational work on trust in automation emphasizes that appropriate trust depends on the relationship between system capabilities, task demands, and the conditions under which reliance occurs \cite{rechkemmer2022confidence, lee2004trust}. In expert settings, these judgments are further complicated by the presence of a human professional \cite{vereschak2024trust} whose competence, responsibility, and appropriate use of technological assistance may also be evaluated. Consequently, patterns of human and AI involvement may be interpreted differently across professional domains.

Recent work provides direct evidence that AI trust can vary across domains. \citet{ploug2026drivers}, for example, compared factors influencing trust in AI across healthcare, finance, and law and found that the relative importance of different trust-relevant attributes varied by application domain. \citet{grankvist2026trust} similarly compared evaluations of supportive and autonomous AI across healthcare, finance, education, and workplace decision making, demonstrating that perceptions of AI roles are meaningfully contextual even when broader preferences recur across domains. Such findings challenge the assumption that results obtained in one application area can automatically be generalized to another.

This contextual variation is consistent with broader theories of expert trustworthiness. Trustworthiness judgments can involve distinct perceptions of ability or expertise, integrity, and benevolence \cite{hendriks2015meti, mayer1995integrative}, and the evidence used to form these judgments may depend on expectations surrounding a particular professional role. Expectations surrounding appropriate consultation, independent judgment, responsible technology use, and competent performance may differ across expert settings \cite{ploug2026drivers, grankvist2026trust}. Domain can consequently influence not only baseline levels of trust, but also the meaning users assign to particular behavioral and procedural cues \cite{ploug2026drivers, schaffernak2026effects}.

Existing cross-domain research shows that trust-related evaluations can vary across applications \cite{grankvist2026trust, ploug2026drivers}, but it does not resolve whether the evaluative consequences of the same observable expert failure generalize when the underlying human-AI workflow is held structurally similar. This question is especially relevant when AI corrects the error and the overall workflow succeeds \cite{kim2026implications}. If users respond primarily to which actor made the error, performance-related recalibration may remain directed toward the human expert across contexts, whereas interpretations of how AI enters the workflow may depend more strongly on domain-specific expectations. 

\subsection{How AI Integration Structure Shapes Evaluations of Human-AI Decision Making}

AI support can be incorporated into expert workflows in structurally different ways. Automation may operate continuously in the background, intervene automatically when predefined conditions are met, or become active only when a human explicitly invokes assistance. Classic work on levels and stages of automation established that these different arrangements change the distribution of human and machine involvement in decision processes \cite{parasuraman2000model}. More recent research suggests that such integration choices can also affect how users evaluate the human professional, the AI system, and the surrounding decision process \cite{shaffer2013derogate, kim2026implications, ackerhans2025perceived, schaffernak2026effects, chen2025impact}. In such workflows, adaptively altering whether and when AI assistance is revealed can promote more appropriate reliance, though often introducing higher perceived task complexity and cognitive demand \cite{ma2023should, sivaraman2023ignore}.

Visible consultation provides one reason why integration structure may become socially meaningful. Advice-seeking research is relevant here not because interpersonal advice seeking is equivalent to AI consultation, but because it shows that observable consultation can itself become evidence from which observers infer characteristics of the decision maker \cite{brooks2015advice}. Related work in medical AI similarly shows that visible physician AI use can change trust in the physician and willingness to seek care \cite{chen2025impact}. Notably, deeper integration of AI-generated advice into the clinical workflow was associated with greater trust in the AI-based decision-support system \cite{ackerhans2025perceived}. \citet{schaffernak2026effects} further showed that the type and timing of clinician-facing AI support can affect evaluations of the physician and the AI-supported decision process. Together, these studies establish that AI integration structure is not merely a technical implementation detail. The visibility, timing, and relationship of AI support to human judgment can become cues through which users interpret both the technology and the professional using it. 

This body of work motivates treating automatic versus expert-initiated AI as a potentially meaningful procedural cue. Because both AI-supported conditions retain the human expert and provide correct AI assistance, their comparison isolates how AI enters the workflow. Prior work does not determine how observers will interpret that difference, whether its meaning varies across domains (RQ3), or whether it changes how expert performance is evaluated (RQ4).

\subsection{Performance, Attribution, and Trust Allocation in Human-AI Systems}

\subsubsection*{Trust Calibration and Expert Performance}

Trust in automation should calibrate to evidence about system capabilities and performance because insensitivity to such evidence can produce inappropriate over- or underreliance \cite{rechkemmer2022confidence, lee2004trust}. In expert settings, observed performance provides evidence relevant to perceived ability or expertise \cite{hendriks2015meti, mayer1995integrative}, and prior work shows that credibility and trust in AI-assisted decision making can change in response to observed performance, error, and recovery \cite{HanKo2025TrustDynamicsFA, pareek2024trust, tenney2008calibration}. Trust-related evaluations should therefore be treated as responsive to interaction evidence rather than as fixed judgments.

\subsubsection*{Expert Failure Versus Final Workflow Success}
In human-AI workflows, AI can correct an expert error, allowing the overall workflow to succeed despite the human's failure. Human-AI workflows can separate expert performance from final workflow success when AI detects or corrects an expert error \cite{kim2026implications}. In the medical study, we hold final workflow success constant while varying observable expert performance, allowing us to test whether human failure remains evaluatively consequential despite successful workflow completion.

\subsubsection*{Attribution of Error and Responsibility}

When different actors contribute to different stages of a workflow, the basis for evaluation becomes more complex. Prior research shows that responsibility and fault in human-AI systems are not necessarily assigned uniformly across contributors \cite{lima2021moral, lima2023blaming}. Observers distinguish between human and AI responsibility for collaborative outcomes, and the structure of joint human-AI decision making can alter how fault is allocated \cite{lima2021moral, shank2019artificial, hu2026responsible}. The form of AI involvement can influence how credit or blame is assigned to human users \cite{earp2024credit}. These findings establish attribution as a relevant problem for mixed human-AI workflows and motivate examining whether the evaluative consequences of an expert error are distributed uniformly across the human, AI, and combined system.

\subsubsection*{Human, AI, and Team as Distinct Trust Targets}

Human-AI systems can contain several conceptually distinct trustees. \citet{riedl2024patients} treat human, AI, and blended human-AI arrangements as distinguishable objects of evaluation. \citet{georganta2024my} further distinguish interpersonal trust in an individual teammate from trust in the team as a collective entity and show that these levels of trust can be analytically separated in human-AI teams. Complementing these distinctions, \citet{vereschak2024trust} show that trust in AI-assisted decision making is shaped by relationships among multiple actors within a broader socio-technical system, rather than by perceptions of the AI or human user in isolation. These approaches caution against collapsing trust across components of a mixed human-AI system.
This distinction is particularly important when one component fails and another supports recovery. The evaluative consequences of an expert error may remain localized to the expert, spill over to the combined human-AI team, or be distributed across multiple targets. In the prior advising study, expert errors produced strong human-directed penalties, whereas AI-directed trust was comparatively insensitive and team-directed effects were weaker and less consistent. More generally, these distinctions motivate testing whether the consequences of an expert error remain localized to the human expert or extend to other components of the human-AI system.

\subsection{Multiple Dimensions of Expert Evaluation and Intended Reuse}

Trust-related evaluation is multidimensional. Treating every outcome as a single generic measure of “trust'' can obscure differences in what users are actually judging. \citet{mayer1995integrative} distinguish ability, integrity, and benevolence as separate bases of perceived trustworthiness, while \citet{hendriks2015meti} developed the Muenster Epistemic Trustworthiness Inventory (METI) to assess expertise, integrity, and benevolence in evaluations of knowledge-based experts. By using METI in our study, we accordingly analyze the three dimensions separately. Expertise captures perceived competence and relevant knowledge, integrity concerns whether the expert is perceived as adhering to appropriate principles or professional standards, and benevolence concerns perceived goodwill toward the person affected by the expert's actions \cite{hendriks2015meti}. 

Alongside METI and target-specific trust \cite{riedl2024patients}, we analyze general trust separately as a broader evaluation of the human expert, allowing global evaluations to be distinguished from more specific judgments of expertise, integrity, and benevolence. Trust can influence subsequent reliance and behavior without being equivalent to them \cite{lee2004trust, liao2022designing}. Accordingly, we treat intended reuse as a distinct behavioral-intention outcome rather than a direct measure of trust. An effect on one outcome therefore should not be interpreted as a change in ``trust as a whole.'' Keeping these outcomes distinct allows us to identify which evaluative patterns generalize across domains and which reveal more specific boundary conditions.

\section{Methodology}

\subsection{Study Design}
The medical study used a 3 $\times$ 2 between-subjects factorial design crossing AI Modality (expert-only, automatic AI, expert-initiated AI) with expert Performance (failure vs. success). Participants were randomly assigned to one of six conditions. The factorial structure and core measures were aligned with those used by \citet{kim2026implications}, enabling subsequent cross-domain harmonization. Apart from replacing the interactive advising interface with a shorter video-based medical simulation and adapting domain-specific terminology, the conceptual structure was preserved.

Participants were asked to imagine themselves as patients undergoing a deep vein thrombosis (DVT) ultrasound scan. The expert's task was to obtain a satisfactory scan that could subsequently be forwarded to a professional diagnostician for interpretation. Participants therefore observed the quality of the technician's scanning procedure rather than receiving or evaluating a diagnosis themselves. Because participants occupied the role of patients without the specialized expertise needed to independently evaluate the technical quality of the scan, the scenario preserved the epistemic-dependence structure of the advising paradigm \cite{kim2026implications}. The interaction was presented as a video simulation showing the male-presenting patient's leg, the male-presenting technician, the ultrasound device, and the ongoing procedure, as well as various scan perspectives, with AI feedback delivered using a male-sounding synthesized voice.

The simulated scanning workflow was consistent with established guidance for lower-extremity venous ultrasound in the evaluation of suspected DVT \cite{needleman2018ultrasound}. The workflow also contained discrete procedural checkpoints and observable scan-quality criteria, allowing technician errors and corresponding corrective feedback to be manipulated consistently across conditions. The scenario, procedural sequence, error manipulations, and AI-provided feedback were also reviewed in consultation with domain experts familiar with DVT ultrasound procedures to support their procedural and clinical plausibility. 

\subsubsection*{AI intervention modality}

The three AI modality conditions differed in how possible problems with the scan were checked and corrected.

\textbf{Expert-only condition.} No AI system was used. At each of two points of suspected error, the technician consulted a reference book containing examples of satisfactory scans. If an error was present, the technician identified and corrected it using the reference material. If no error was present, the technician confirmed that the scan was satisfactory and continued the procedure.

\textbf{Expert-initiated AI condition.} AI assistance was available but had to be explicitly invoked by the technician. At each point of suspected error, the technician initiated an AI check. When an error was present, the AI identified the problem and provided corrective guidance, which the technician followed before continuing. When no error was present, the AI confirmed that the scan was satisfactory. 

\textbf{Automatic AI condition.} The AI continuously monitored the procedure without requiring technician invocation. At each checkpoint, the AI automatically either confirmed satisfactory performance or identified an error. When an error was identified, the technician corrected the scan in accordance with the AI's instruction before continuing.

Figure~\ref{fig:medical_workflows} illustrates how the three workflow structures were instantiated in the medical simulation.

\subsubsection*{Technician performance}

Technician performance was manipulated independently of AI modality. In the failure condition, the technician made errors at both designated checkpoints. These errors involved either losing the vein during the scan or producing an image of insufficient quality. Each error was subsequently detected through the modality-specific checking mechanism and corrected by the technician before the procedure continued.

In the success condition, the technician completed the scan without making either of the two possible errors. At each checkpoint, the applicable verification mechanism, the reference book, expert-initiated AI, or automatic AI, confirmed satisfactory performance. There were no mixed-performance conditions: participants in the failure condition observed errors at both checkpoints, whereas participants in the success condition observed no technician errors.

Importantly, the final scan was successfully completed in every condition. AI correctness was held fixed by design: when AI was present, its judgments and corrective guidance were always consistent with the intended procedure and had been reviewed with domain experts, so participants were not exposed to AI errors. The manipulation therefore separated observable technician performance from final workflow success: a participant could observe the technician make and correct an error while still ultimately observing a successful scan. 

\subsection{Participants}

Participants were recruited through Prolific \cite{prolific} and completed the study remotely using a desktop or laptop computer. Repeat participation was prevented using Prolific's eligibility controls. To enable a balanced cross-domain comparison, we targeted a sample size comparable to that of the advising study ($n = 157$) \cite{kim2026implications}. A total of 170 participants completed the study; four were excluded because of failed attention checks or poor-quality responses, yielding a final medical sample of $N = 166$. Participants received \$4.50 and completed the study in approximately 14 minutes on average. The study procedures were approved by the institutional review board of the authors' institution, and informed consent was obtained from all participants. Detailed participant demographic information is reported in the Appendix (Table~\ref{tab:participant_demographics_wide}).

\subsection{Measures}

Outcome measures targeted three levels of evaluation: multidimensional epistemic trustworthiness of the technician; entity-specific trust in the technician, the AI assistant, and their team; and a global trust rating and intended reuse.

\subsubsection*{Epistemic trustworthiness}

Participants completed items from the Muenster Epistemic Trustworthiness Inventory \cite{hendriks2015meti}. Similar to \citet{hendriks2015meti}, items were aggregated separately into expertise, integrity, and benevolence dimensions. Higher scores indicate greater perceived epistemic trustworthiness. In the medical sample, internal consistency was high for expertise (7 items, Cronbach's $\alpha = .959$), integrity (5 items, $\alpha = .870$), and benevolence (4 items, $\alpha = .891$). The three METI dimensions were analyzed separately rather than combined into a single trustworthiness score.

\subsubsection*{Entity-specific trust}

Entity-specific trust was assessed using the parallel item structure adapted from \citet{riedl2024patients}. Participants rated the technician on three items assessing perceived trustworthiness, affective comfort when relying on the technician, and trust in the information presented by the technician. Participants in AI-supported conditions additionally completed the same three items for the AI assistant and for the technician-AI combination. Wording and response scales were unchanged from the advising version \cite{kim2026implications}, apart from substituting the relevant medical-domain actors. In the medical sample, the three-item blocks showed high internal consistency for the human ($\alpha = .945$), AI ($\alpha = .958$), and human--AI team ($\alpha = .963$) targets.  Despite high internal consistency, each item was evaluated individually to avoid collapsing conceptually distinct dimensions of entity-specific evaluation.

\subsubsection*{General trust}

Participants provided a global rating of the technician in response to: ``From 1-100, how trustworthy would you rate your technician? (1 being the least trustworthy and 100 being the most).'' The 1–100 response scale was retained unchanged from the advising study to preserve measurement comparability across domains. This provided a single-item global benchmark to capture overall trust in the professional alongside granular multidimensional ratings.

\subsubsection*{Intended reuse}

Participants reported how likely they would be to use the technician again on a 7-point Likert-type scale ranging from 1 (\emph{extremely unlikely}) to 7 (\emph{extremely likely}). Participants in AI-supported medical conditions additionally rated the likelihood of reusing the AI assistant and the technician-AI combination. For the cross-domain analyses, only human-expert reuse was analyzed because the advising study \cite{kim2026implications} did not administer corresponding AI- or team-reuse items. Reuse is interpreted as a behavioral-intention outcome rather than as a direct measure of trust.

\subsubsection*{Free Responses}

Participants also had opportunities to provide optional free-text comments about their evaluations and the observed workflow.

\subsection{Procedure}

The survey was built in Qualtrics \cite{qualtrics}. After providing informed consent, participants received instructions introducing the medical scenario and their role as the patient. Qualtrics randomly assigned each participant to one of the six experimental conditions. Participants then viewed the video simulation corresponding to their assigned technician-performance and AI-modality condition. Following the simulation, participants completed an attention-check question, the trust and reuse measures, and demographic questions before receiving a payment-completion code.

To ensure data quality, the attention check followed the simulation and assessed whether participants had attended to relevant features of the workflow. Participants in the failure conditions were asked about the nature of the observed failure points, while those in the success conditions were asked about a corresponding detail of the simulation. Participants who failed the attention check or did not meet the quality criteria were excluded from the analysis.

\subsection{Cross-Domain Harmonization and Analysis}

To examine cross-domain generalizability, we obtained the pre-existing academic-advising data from the authors of \citet{kim2026implications} and harmonized them with our medical dataset. The resulting combined sample included 157 advising participants and 166 medical participants (\(N=323\)). Variables were renamed to common labels where necessary. Our questionnaire included two additional reuse questions with no advising-domain analogues - these medical-only variables were excluded from the cross-domain analysis. Performance was harmonized as failure versus success, and modality was harmonized as expert-only, automatic AI, and expert-initiated AI.

For outcomes available under all modality conditions, we conducted 2 $\times$ 3 $\times$ 2 Type II factorial ANOVAs crossing Domain (advising vs.\ medicine), AI Modality (expert-only, automatic AI, expert-initiated AI), and Performance (failure vs.\ success), $N = 323$. Outcomes were METI expertise, integrity, and benevolence; the three human-directed Riedl items; general trust; and human-expert reuse. Because AI- and team-directed ratings were administered only when AI was present, a second analysis was restricted to the automatic and expert-initiated AI conditions ($N = 219$). These 2 $\times$ 2 $\times$ 2 Type II ANOVAs crossed Domain, AI Modality, and Performance and included all nine Riedl items in addition to the shared human-facing outcomes. This restricted model also provided a direct cross-domain comparison of automatic versus expert-initiated AI without the expert-only condition. The factorial ANOVAs constituted the primary analyses, and effect sizes for the combined-domain models are reported as partial $\eta^2$.

Because multiple conceptually distinct outcomes were analyzed, we controlled multiplicity within predefined outcome families for each focal factorial effect using Benjamini-Hochberg false-discovery-rate (BH-FDR) correction, with Holm correction as a more conservative companion analysis. Full outcome-family definitions and correction details are reported in Appendix~\ref{app:multiplicity}.

Analyses were conducted in Python. Residuals and variance assumptions were examined using residual diagnostics, Levene tests, and Levene/Brown-Forsythe tests where appropriate. Because several outcomes showed evidence of unequal variances, we additionally conducted HC3 heteroskedasticity-robust sensitivity analyses. HC3 covariance estimates were used to construct robust Wald tests corresponding to the factorial terms. These tests are related to, but not identical to, the conventional Type II ANOVA tests and were therefore used to assess whether substantive conclusions were sensitive to heteroskedasticity rather than as replacements for the primary ANOVAs.

For the Domain $\times$ Modality interaction on perceived expertise, targeted exploratory decompositions used model-based estimated marginal means and pairwise modality contrasts within each domain, averaging equally over Performance. Multiplicity was controlled across the corresponding contrasts using Holm adjustments. Exploratory Tukey HSD comparisons across factorial cells were retained for continuity with the prior analysis but were not used as the primary basis for cross-domain inference.

Human-expert reuse was measured on a bounded 1-7 ordinal scale and showed sensitivity to inferential specification. We therefore conducted additional exploratory sensitivity analyses using cumulative-link ordinal models. For both the full and AI-supported analyses, cumulative-logit models were fit hierarchically, and the Domain $\times$ Modality $\times$ Performance interaction was evaluated using a likelihood-ratio test comparing the full model with a nested model omitting the three-way interaction while retaining the corresponding lower-order terms. All ordinal models used in the final analyses converged successfully. Cumulative-probit models were additionally fit as a link-function sensitivity check. Agreement between logit and probit specifications assesses sensitivity to the choice of link function - it does not test the common-effect assumption across response thresholds. The cumulative-link models assume common predictor effects across thresholds.

To characterize the reuse interaction, we decomposed the full ordinal-logit model into all 12 modality contrasts within fixed Domain $\times$ Performance cells: three modality comparisons within each of the four Domain $\times$ Performance combinations. Wald contrasts were derived from the fitted ordinal model, with Holm correction applied once across all 12 comparisons. The same 12 comparisons were additionally evaluated using HC3-based linear contrasts and two-sided Mann-Whitney tests as convergent sensitivity checks, with Holm correction applied separately across the 12 contrasts within each method.

The factorial ANOVAs remained the primary basis for inference. Multiplicity corrections, HC3 robust analyses, ordinal reuse models, and targeted follow-up decompositions were diagnostic- and robustness-motivated analyses rather than preregistered confirmatory tests.

\section{Results}

Throughout, we report Type II ANOVA $F$-tests with partial eta squared ($\etap$) as the effect size. We first characterize the collected medical study using two-way (Modality $\times$ Performance) ANOVAs. We then use the combined-domain models to test whether the observed effects vary between medicine and academic advising. The combined-domain analyses are three-way ANOVAs that add Domain as a factor, either with all three modality levels (the full 2 $\times$ 3 $\times$ 2 model) or restricted to the two AI-supported modalities (the 2 $\times$ 2 $\times$ 2 AI-supported model). Cross-domain inferences are based on these models containing Domain directly as a factor, rather than on comparisons of statistical significance across separate advising and medical analyses. Descriptive statistics for all outcomes by condition are reported in Table~\ref{tab:descriptives}.

\subsection{Evaluation of AI Modality and Performance in a High-Stakes Domain}

We first analyzed the collected medical data using 3 $\times$ 2 ANOVAs. AI Modality significantly affected expertise, general trust, technician reuse, affective comfort toward the technician, and trust in technician-provided information (Table~\ref{tab:medical_modality_anova}). Modality did not significantly affect integrity, benevolence, or perceived trustworthiness of the technician. Descriptively, the automatic-AI condition produced the highest mean ratings on each of the five outcomes showing a significant modality effect, with particularly clear separations for expertise, intended reuse, and affective comfort. Success-condition means were also higher than failure-condition means across all eight human-facing outcomes, a performance pattern examined inferentially under H1 in the following section. Because differences in statistical significance across separate medical and advising analyses \cite{kim2026implications} do not establish a domain difference, we test Domain directly in the combined analyses below.

\begin{table}[htbp]
\caption{Medical-domain AI Modality effects from the 3 $\times$ 2 Type II ANOVAs ($N = 166$).}
\Description{Table reporting the main effects of AI Modality on eight medical-domain outcomes. Significant modality effects were found for perceived expertise, intended reuse, general trust, affective comfort toward the technician, and trust in technician-provided information. Modality effects were not significant for integrity, benevolence, or perceived trustworthiness of the technician. The table reports F statistics, p-values, and partial eta-squared effect sizes for each outcome.}
\label{tab:medical_modality_anova}
\centering
\small
\begin{tabular}{lccc}
\toprule
Outcome & $F(2,160)$ & $p$ & $\eta_p^2$ \\
\midrule
Expertise                & 9.17 & $<.001$ & .10 \\
Integrity                & 2.57 & .080    & .03 \\
Benevolence              & 0.86 & .425    & .01 \\
Intended reuse           & 7.06 & .001    & .08 \\
General trust            & 4.91 & .009    & .06 \\
Trustworthiness (Tech)   & 2.95 & .055    & .04 \\
Affective comfort (Tech)  & 8.24 & $<.001$ & .09 \\
Information trust (Tech) & 4.99 & .008    & .06 \\
\bottomrule
\end{tabular}
\end{table}


\subsection{Expert Errors Reduce Evaluations of the Human Expert (H1)}

H1 was supported: in the full 2 $\times$ 3 $\times$ 2 model, expert failure lowered all eight shared human-facing outcomes, and all eight Performance effects remained significant after both BH-FDR and Holm correction (Table~\ref{tab:full_performance_effects}). Higher-order interactions involving performance, modality, and domain are reported in Table~\ref{tab:domain_interactions_anova232}. These results indicate a broad human-facing evaluative penalty for expert error.

\begin{table}[htbp]
\centering
\small
\setlength{\tabcolsep}{4pt}
\caption{Domain-involving interaction effects from the $2\times3\times2$ Type II ANOVAs ($N=323$). D = Domain, M = Modality, P = Performance, Trustworthiness (expert) = The technician is trustworthy, Affective comfort (expert) = I have a good feeling when relying on the technician, Information trust (expert) = I can trust the information presented by the technician.}
\Description{Table reporting Domain-by-Modality, Domain-by-Performance, and Domain-by-Modality-by-Performance interaction effects for eight outcomes in the combined advising and medical sample. The only statistically significant interaction was Domain by Modality for perceived expertise. No Domain-by-Performance interactions were significant. The Domain-by-Modality-by-Performance interaction for reuse intention approached but did not reach the conventional significance threshold. The table reports F statistics, p-values, and partial eta-squared effect sizes for each interaction and outcome.}

\label{tab:domain_interactions_anova232}
\begin{tabular}{l ccc ccc ccc}
\toprule
 & \multicolumn{3}{c}{D$\times$M, $F(2,311)$} & \multicolumn{3}{c}{D$\times$P, $F(1,311)$} & \multicolumn{3}{c}{D$\times$M$\times$P, $F(2,311)$}\\
\cmidrule(lr){2-4}\cmidrule(lr){5-7}\cmidrule(lr){8-10}
Outcome & $F$ & $p$ & $\eta^2_p$ & $F$ & $p$ & $\eta^2_p$ & $F$ & $p$ & $\eta^2_p$\\
\midrule
Expertise & 6.39 & .002 & .039 & 0.19 & .663 & .001 & 1.05 & .352 & .007\\
Integrity & 1.09 & .336 & .007 & 0.00 & .989 & .000 & 0.33 & .721 & .002\\
Benevolence & 0.57 & .567 & .004 & 1.11 & .294 & .004 & 0.09 & .914 & .001\\
Reuse intention & 2.81 & .062 & .018 & 1.68 & .196 & .005 & 2.95 & .054 & .019\\
General trust & 1.66 & .192 & .011 & 3.02 & .083 & .010 & 1.42 & .243 & .009\\
Trustworthiness (expert) & 1.14 & .320 & .007 & 1.06 & .304 & .003 & 0.31 & .731 & .002\\
Affective comfort (expert) & 1.79 & .168 & .011 & 1.90 & .169 & .006 & 0.65 & .522 & .004\\
Information trust (expert) & 1.92 & .148 & .012 & 2.13 & .145 & .007 & 0.23 & .796 & .001\\
\bottomrule
\end{tabular}
\end{table}

\subsection{Trust Consequences Are Asymmetrically Distributed Across Human, AI, and Team Targets (H2, RQ1)}

H2 and RQ1 were evaluated in the AI-supported subset ($N = 219$). Performance significantly affected all three human-directed trust items, whereas none of the AI-directed or team-directed performance effects reached significance (Table~\ref{tab:target_specific_performance}). This human-sensitive, AI-insensitive pattern is consistent with H2. Because we did not estimate a Target $\times$ Performance interaction, however, it does not establish that the Performance effect was statistically larger for the human than for the AI.

Within-domain team patterns differed descriptively: \citet{kim2026implications} reported significant team-directed Performance effects on two of the three outcomes in advising, whereas none were significant in medicine. However, all three Domain $\times$ Performance interactions for team-directed outcomes were nonsignificant (Table~\ref{tab:target_specific_performance}), providing no evidence that team spillover differed systematically by domain. Thus, expert failure reliably lowered human-directed trust and left AI-directed trust comparatively insensitive, while evidence of team-directed spillover was weaker and did not differ detectably across domains.

\subsection{Performance-Related Trust Effects Show Little Evidence of Domain Moderation (RQ2)}

For RQ2, we found no evidence of domain moderation: none of the eight shared human-facing Domain $\times$ Performance interactions was significant (Table~\ref{tab:domain_interactions_anova232}). These nonsignificant interactions do not establish statistical equivalence, but they provide no evidence that the performance penalty differed systematically between advising and medicine.

\subsection{AI Modality Shows Selective Domain Dependence (RQ3)}

\begin{figure}[!t]
    \centering
    \includegraphics[width=\columnwidth]
    {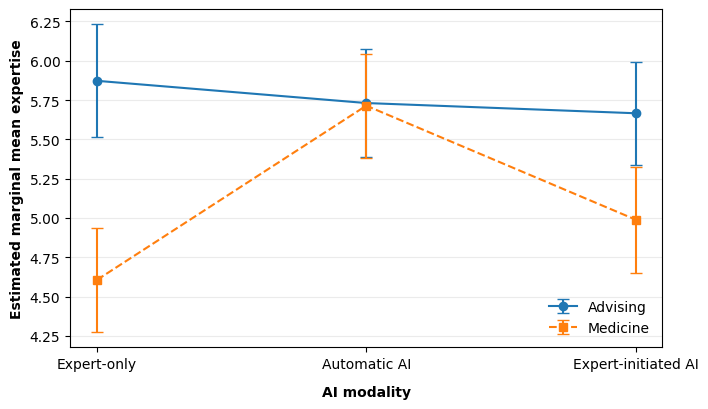}
    \caption{Domain $\times$ Modality interaction for perceived expertise in the full $2\times3\times2$ model. Points show estimated marginal means averaging equally over Performance; error bars show model-based 95\% confidence intervals.}
    \Description{Line plot of estimated marginal mean expertise across AI modalities for the advising and medical domains. Expertise ratings remain relatively stable across modalities in advising. In medicine, ratings are lowest in the human-only condition, increase under automatic AI, and decrease under expert-initiated AI. Error bars show 95\% confidence intervals.}
    \label{fig:expertise_domain_modality}
\end{figure}

Perceived expertise provided the clearest evidence for RQ3: the full-model Domain $\times$ Modality interaction was significant (Table~\ref{tab:domain_interactions_anova232}) and remained significant after both BH-FDR and Holm correction across the eight shared outcomes (adjusted $p = .0153$ under both procedures). Follow-up decomposition showed no Holm-significant modality separation in advising. In medicine, automatic AI was associated with higher perceived expertise than both expert-only and expert-initiated AI, whereas expert-initiated AI did not differ detectably from expert-only (Table~\ref{tab:expertise_decomposition}; Figure~\ref{fig:expertise_domain_modality}). The medical pattern is therefore best characterized as an automatic-oversight elevation rather than an expert-initiated AI penalty.

\begin{table}[!t]
\centering
\small
\caption{Exploratory decomposition of the Domain $\times$ Modality interaction
for perceived expertise in the full $2\times3\times2$ model. Estimated marginal
means average equally over Performance. For contrasts, $\Delta M$ is the
first-listed modality minus the second-listed modality. Holm adjustment was
applied across the six within-domain modality contrasts.}
\label{tab:expertise_decomposition}
\Description{Two-panel table decomposing the Domain-by-Modality interaction for perceived expertise. Panel A reports estimated marginal means and 95\% confidence intervals for expert-only, automatic AI, and expert-initiated AI conditions within advising and medicine. The three modalities showed similar expertise ratings in advising. In medicine, automatic AI had higher estimated expertise than both expert-only and expert-initiated AI. Panel B reports within-domain pairwise modality contrasts with Holm-adjusted p-values. No advising-domain contrasts were significant. In medicine, automatic AI was significantly higher than expert-only and expert-initiated AI, while expert-only and expert-initiated AI did not differ significantly.}

\textit{Panel A: Estimated marginal means}\\[2pt]

\begin{tabular}{llcc}
\toprule
Domain & Modality & EMM & 95\% CI \\
\midrule
Advising & Expert-only          & 5.87 & [5.51, 6.23] \\
Advising & Automatic AI        & 5.73 & [5.39, 6.08] \\
Advising & Expert-initiated AI & 5.67 & [5.34, 5.99] \\
Medicine & Expert-only          & 4.61 & [4.27, 4.94] \\
Medicine & Automatic AI        & 5.71 & [5.38, 6.05] \\
Medicine & Expert-initiated AI & 4.99 & [4.65, 5.33] \\
\bottomrule
\end{tabular}

\vspace{6pt}

\textit{Panel B: Within-domain modality contrasts}\\[2pt]

\begin{tabular}{llccc}
\toprule
Domain & Comparison & $\Delta M$ & 95\% CI & $p_{\mathrm{Holm}}$ \\
\midrule
Advising
& Expert-only $-$ Automatic AI
& 0.14 & [-0.36, 0.64] & 1.000 \\

Advising
& Expert-only $-$ Expert-initiated AI
& 0.21 & [-0.28, 0.69] & 1.000 \\

Advising
& Automatic AI $-$ Expert-initiated AI
& 0.07 & [-0.41, 0.54] & 1.000 \\

Medicine
& Expert-only $-$ Automatic AI
& -1.11 & [-1.58, -0.64] & $<.001$ \\

Medicine
& Expert-only $-$ Expert-initiated AI
& -0.38 & [-0.86, 0.09] & .446 \\

Medicine
& Automatic AI $-$ Expert-initiated AI
& 0.72 & [0.25, 1.20] & .014 \\
\bottomrule
\end{tabular}
\end{table}

The corresponding Domain $\times$ Modality interaction in the AI-supported subset was nominally significant but did not survive across-outcome correction (Table~\ref{tab:ai_supported_interactions_anova222}), so the full three-modality interaction remains the primary evidence for RQ3. No other shared outcome showed multiplicity-robust Domain $\times$ Modality evidence. RQ3 therefore indicates selective domain dependence, most clearly for expertise, without identifying why the difference occurred.

\begin{table*}[!t]
\centering
\scriptsize
\setlength{\tabcolsep}{2.8pt}
\caption{Domain-involving interaction effects from the AI-supported $2\times2\times2$ Type II ANOVAs ($N=219$). All tests have $F(1,211)$. BH-FDR and Holm adjustments were applied within the corresponding predefined outcome family.}
\Description{Table reporting Domain-by-Modality, Domain-by-Performance, and Domain-by-Modality-by-Performance interaction effects for 14 outcomes in the AI-supported subset. For each interaction, the table reports F statistics, unadjusted p-values, partial eta-squared effect sizes, and BH-FDR- and Holm-adjusted p-values. The Domain-by-Modality interactions for perceived expertise and intended reuse were nominally significant before correction, but neither survived BH-FDR or Holm adjustment. No Domain-by-Performance or three-way interactions were statistically significant after correction.}
\label{tab:ai_supported_interactions_anova222}
\begin{tabular}{l ccccc ccccc ccccc}
\toprule
 & \multicolumn{5}{c}{D$\times$M, $F(1,211)$} & \multicolumn{5}{c}{D$\times$P, $F(1,211)$} & \multicolumn{5}{c}{D$\times$M$\times$P, $F(1,211)$}\\
\cmidrule(lr){2-6}\cmidrule(lr){7-11}\cmidrule(lr){12-16}
Outcome & $F$ & $p$ & $\eta^2_p$ & $p_{\mathrm{BH}}$ & $p_{\mathrm{Holm}}$ & $F$ & $p$ & $\eta^2_p$ & $p_{\mathrm{BH}}$ & $p_{\mathrm{Holm}}$ & $F$ & $p$ & $\eta^2_p$ & $p_{\mathrm{BH}}$ & $p_{\mathrm{Holm}}$\\
\midrule
Expertise                  & 4.22 & .041 & .020 & .165 & .288 & 0.16 & .693 & .001 & .792 & 1.000 & 0.46 & .500 & .002 & .800 & 1.000\\
Integrity                  & 1.84 & .177 & .009 & .212 & .663 & 0.22 & .638 & .001 & .792 & 1.000 & 0.00 & .945 & .000 & .945 & 1.000\\
Benevolence                & 0.92 & .338 & .004 & .338 & .663 & 1.24 & .267 & .006 & .729 & 1.000 & 0.03 & .873 & .000 & .945 & 1.000\\
Intended reuse             & 4.72 & .031 & .022 & .165 & .247 & 0.01 & .933 & .000 & .933 & 1.000 & 3.10 & .080 & .014 & .633 & .638\\
General trust              & 2.60 & .109 & .012 & .212 & .543 & 0.83 & .364 & .004 & .729 & 1.000 & 2.00 & .158 & .009 & .633 & 1.000\\
Trustworthiness (AI)       & 1.14 & .286 & .005 & .825 & .859 & 3.04 & .083 & .014 & .124 & .167 & 0.15 & .698 & .001 & .929 & 1.000\\
Affective comfort (AI)      & 0.01 & .926 & .000 & .926 & 1.000 & 2.37 & .125 & .011 & .125 & .167 & 1.58 & .210 & .007 & .629 & .629\\
Information trust (AI)     & 0.36 & .550 & .002 & .825 & 1.000 & 3.71 & .056 & .017 & .124 & .167 & 0.01 & .929 & .000 & .929 & 1.000\\
Trustworthiness (team)     & 0.96 & .328 & .005 & .361 & .655 & 1.39 & .240 & .007 & .360 & .480 & 1.17 & .281 & .006 & .782 & .844\\
Affective comfort (team)    & 0.84 & .361 & .004 & .361 & .655 & 0.43 & .515 & .002 & .515 & .515 & 0.04 & .838 & .000 & .838 & 1.000\\
Information trust (team)   & 2.23 & .136 & .010 & .361 & .409 & 2.25 & .135 & .011 & .360 & .404 & 0.41 & .522 & .002 & .782 & 1.000\\
Trustworthiness (expert)   & 1.93 & .166 & .009 & .212 & .663 & 0.86 & .355 & .004 & .729 & 1.000 & 0.70 & .404 & .003 & .800 & 1.000\\
Affective comfort (expert)  & 3.03 & .083 & .014 & .212 & .500 & 0.29 & .591 & .001 & .792 & 1.000 & 0.03 & .852 & .000 & .945 & 1.000\\
Information trust (expert) & 1.77 & .185 & .008 & .212 & .663 & 1.80 & .182 & .008 & .729 & 1.000 & 0.52 & .470 & .002 & .800 & 1.000\\
\bottomrule
\end{tabular}
\end{table*}

\subsection{Three-Way Moderation Was Generally Limited, with a Reuse Exception (RQ4)}

RQ4 received little support under the primary ANOVAs: no Domain $\times$ Modality $\times$ Performance interaction reached the conventional significance threshold in the full model (Table~\ref{tab:domain_interactions_anova232}), with human-expert reuse providing the closest result. The corresponding reuse interaction was also nonsignificant in the AI-supported model (Table~\ref{tab:ai_supported_interactions_anova222}).

Reuse was nevertheless sensitive to model specification. The full-model HC3 analysis crossed the conventional significance threshold, whereas the AI-supported HC3 analysis did not. Cumulative-logit and cumulative-probit models supported the three-way interaction in both analyses (Table~\ref{tab:reuse_sensitivity}). We therefore treat the reuse interaction as an exploratory sensitivity finding rather than a primary confirmatory result.

\begin{table}[!t]
\centering
\small
\setlength{\tabcolsep}{4pt}
\caption{Exploratory sensitivity analyses for the Domain $\times$ Modality
$\times$ Performance interaction on human-expert intended reuse.
Full = full $2\times3\times2$ model; AI = AI-supported $2\times2\times2$
model. The cumulative-link analyses are exploratory sensitivity analyses,
not replacements for the primary factorial ANOVAs. All ordinal models underlying the reported likelihood-ratio tests converged successfully.}
\Description{Table reporting exploratory sensitivity analyses for the Domain-by-Modality-by-Performance interaction on human-expert intended reuse. Results are shown for the full three-modality model and the AI-supported two-modality model using HC3 robust Wald tests and cumulative-link ordinal models with logit and probit links. In the full model, the HC3 robust test and both cumulative-link models were statistically significant. In the AI-supported model, the HC3 robust test was not significant, while both cumulative-link models were significant.}
\label{tab:reuse_sensitivity}

\begin{tabular}{llcc}
\toprule
Analysis & Test & Statistic & $p$ \\
\midrule
Full & HC3 robust Wald
     & $F(2,311)=3.15$ & .044 \\

AI & HC3 robust Wald
   & $F(1,211)=2.96$ & .087 \\

Full & Cumulative-logit LR
     & $\chi^2(2)=10.79$ & .00455 \\

AI & Cumulative-logit LR
   & $\chi^2(1)=5.10$ & .02395 \\

Full & Cumulative-probit LR
     & $\chi^2(2)=9.83$ & .00733 \\

AI & Cumulative-probit LR
   & $\chi^2(1)=5.75$ & .01649 \\
\bottomrule
\end{tabular}
\end{table}

\begin{figure}[!t]
    \centering
    \includegraphics[width=\columnwidth]{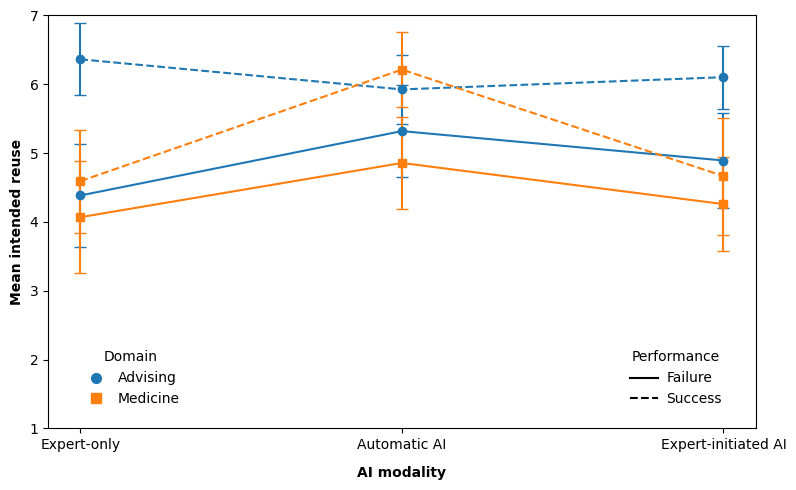}
    \caption{Human-expert intended reuse across Domain, AI Modality, and
Performance. Points show cell means and error bars show 95\% confidence
intervals. The largest modality separation occurs following successful
medical performance, where automatic AI received higher reuse ratings than
both expert-only and expert-initiated AI. Inferential results for the
within-cell modality contrasts are reported in
Table~\ref{tab:reuse_contrasts}.}
\label{fig:reuse_interaction}
\Description{Line plot of mean intended reuse across domain, AI modality, and expert performance. Successful performance generally produces higher reuse ratings than failed performance. In the medical domain, reuse following successful performance is substantially higher with automatic AI than with either human-only or expert-initiated AI. Error bars show 95\% confidence intervals.}

\end{figure}

Fixed-cell contrasts localized the pattern to successful medical performance: automatic AI produced greater intended reuse than both expert-only and expert-initiated AI, while expert-initiated AI did not differ detectably from expert-only (Table~\ref{tab:reuse_contrasts}). The same two contrasts survived Holm correction under ordinal Wald, HC3 linear, and Mann-Whitney analyses, while the other ten did not. Thus, the exploratory evidence indicates a performance-contingent automatic-oversight benefit specific to successful medical performance (Figure~\ref{fig:reuse_interaction}). The expert-initiated AI success cell was unusually dispersed (Table~\ref{tab:reuse_contrasts}), so its mean should not be treated as fully characterizing participant responses.

\subsection{Exploratory Qualitative Observations}

A small number of participants provided free-text comments that offered additional context for interpreting the modality-related patterns. Given the limited number of responses, we did not conduct a formal qualitative analysis or thematic coding. Instead, we used LLM assistance to help surface potentially relevant comments, which were then examined against the original responses. These observations are therefore treated as anecdotal and exploratory rather than as estimates of theme prevalence or evidence of explanatory mechanisms.

Several comments suggested that expert-initiated AI could be interpreted as signaling uncertainty, inexperience, or dependence on assistance. In contrast, some comments characterized automatic AI support as a safeguard or accuracy check. Expert-initiated consultation was not uniformly interpreted negatively. Some comments instead described deliberate checking as cautious, transparent, or conscientious. These observations provide possible interpretations of the quantitative modality effects but do not establish that these perceptions caused them.

\section{Discussion}

Across advising and medicine, three patterns emerged. First, expert failure consistently reduced all eight shared human-facing outcomes, with no evidence that these penalties differed systematically by domain. Second, perceived expertise showed the clearest multiplicity-robust, domain-sensitive modality effect: in medicine, automatic AI oversight produced higher ratings than both expert-only and expert-initiated AI, while expert-initiated AI did not differ detectably from expert-only and no corresponding modality separation was supported in advising. Third, intended reuse showed a similar three-modality ordering only after successful medical performance. Although the primary three-way ANOVA did not reach the conventional significance threshold, exploratory sensitivity analyses supported this pattern. \textbf{Together, these findings distinguish a comparatively portable response to expert performance from more selective, context-sensitive responses to AI integration structure.} Because expertise and reuse differed in interaction structure and evidential status, we do not assume that they reflect the same psychological mechanism.

\subsection{Cross-Domain Stability of Performance-Related Trust}

\textbf{Expert failure remained consequential even when AI or other checking preserved a successful final outcome.} Across both domains, the broad human-facing penalty is consistent with trust updating to the expert's own observed performance rather than only the workflow endpoint \cite{lee2004trust, tenney2008calibration, pareek2024trust}. The nonsignificant Domain $\times$ Performance interactions do not establish equivalence, but they provide no evidence that this recalibration differed systematically between advising and medicine. Within the two domains studied, it appears comparatively portable.

Target-specific results further suggest localization toward the observed source of failure: human-directed trust changed with expert performance, AI-directed trust remained comparatively insensitive, and team-directed Performance effects were not consistently detected, with no evidence of cross-domain moderation. Because we did not test Target $\times$ Performance or directly measure responsibility attribution, we cannot claim a statistically larger human effect or a causal attribution mechanism. The pattern is nevertheless consistent with prior work showing that responsibility, fault, credit, and blame can be distributed unevenly across contributors to human-AI systems \cite{lima2023blaming, earp2024credit, hu2026responsible, shank2019artificial}.

\subsection{Domain Sensitivity of AI Integration Structure}

The clearest domain-sensitive modality finding was an automatic-oversight elevation in medical expertise, not a detectable penalty for expert-initiated AI.

This pattern does not align straightforwardly with either of two prior findings that suggest different consequences of visible assistance. \citet{shaffer2013derogate} found that patients evaluated physicians less favorably when they used computerized diagnostic support, which could suggest a penalty for visible technological assistance. In contrast, \citet{brooks2015advice} found that seeking advice could increase perceived competence, which could suggest a benefit for expert-initiated consultation. We observed neither pattern directly. Expert-initiated AI did not differ detectably from expert-only, while automatic oversight produced higher perceived expertise than both. This suggests that evaluations may depend not simply on whether assistance is used or sought, but on how that assistance enters the expert workflow.

Qualitative responses suggest one possible interpretation: expert-initiated AI was sometimes read as uncertainty, inexperience, or dependence, whereas automatic support was described as a safeguard or accuracy check. This interpretation is consistent with prior work showing that visible professional AI use and the structure or timing of AI assistance can shape evaluations \cite{chen2025impact, ackerhans2025perceived, schaffernak2026effects}. Yet expert-initiated AI was not interpreted uniformly negatively - some participants viewed deliberate consultation as cautious, transparent, or conscientious. It may therefore function as an ambiguous social signal. The dispersed reuse responses in the expert-initiated AI success condition are consistent with this heterogeneity but do not establish that these interpretations caused the quantitative pattern.

Perceived expertise and intended reuse converged in their ordering but differed in interaction structure. In medicine, both favored automatic oversight over the other workflows in the relevant comparisons, but expertise showed no further moderation by Performance, whereas the reuse ordering appeared only after successful performance. One possibility is that when no visible error makes assistance obviously necessary, observers may rely more on the procedural meaning of AI involvement when forming reuse intentions. This remains a hypothesis rather than a causal conclusion. These findings should not be taken as evidence that expertise and reuse arise from a common psychological mechanism, that automatic AI is generally preferable, or that medical risk or stakes caused the observed differences.

\subsection{Multidimensional Evaluation and Design Implications}

These results reinforce the value of treating expert evaluation as multidimensional. In the cross-domain analysis, Performance affected expertise, integrity, and benevolence, but modality's clearest cross-domain effect was on expertise. The differing sensitivity of these outcomes is consistent with multidimensional accounts of trustworthiness \cite{hendriks2015meti,mayer1995integrative}. Observable mistakes and procedural errors can inform perceptions of capability and professional standards, whereas AI integration structure may communicate competence-related information without providing corresponding evidence about integrity or benevolent intent. The absence of comparably strong cross-domain modality evidence for general trust likewise suggests that a workflow can change perceived capability without producing an equally broad shift in overall trust. Keeping these outcomes separate therefore reveals boundary conditions that a single composite trust measure could obscure.

How AI enters the workflow is also a socially meaningful design feature, not only a technical one. Our findings suggest that designers should consider not only whether AI assistance is available, but also how its initiation is presented to users. In settings where continuous AI oversight is appropriate, automatic monitoring may allow AI to function as a routine safeguard without making each instance of assistance a visible signal of expert uncertainty. Conversely, when expert initiation is necessary, interfaces could frame consultation as routine verification or quality assurance rather than leaving users to infer why the expert sought assistance. These recommendations do not imply that automatic oversight is universally preferable. Rather, designers should consider the social meaning communicated by the integration structure alongside accuracy, timing, usability, and professional norms.

Our qualitative responses also point to a more specific design possibility. One participant suggested that, rather than visibly asking the AI whether the scan was correct, the technician could receive AI alerts through the interface and present the process as a routine completion check. Although this single response is anecdotal, it illustrates how the same corrective functionality could be retained while making the expert--AI interaction mechanics less salient to the patient. This possibility is consistent with the broader pattern in the free responses, where automatic AI was sometimes interpreted as routine quality control or a safeguard, while explicit invocation could signal uncertainty or dependence.

A related design question concerns whether users need to observe the precise interaction mechanics between the expert and AI. In some medical settings, it may be appropriate to disclose that AI contributes to a workflow without necessarily exposing whether a particular AI check was automatically triggered or explicitly requested by the clinician. Existing medical-ethics work rejects a universal requirement that every use of medical machine learning must be disclosed to patients \cite{hatherley2025disclosure}, while other work proposes different levels of patient notification or informed consent depending on factors including AI autonomy, departure from standard practice, patient-facing involvement, and clinical risk \cite{rose2024notification}. These arguments do not justify concealing materially relevant AI involvement. Instead, they suggest that disclosure of AI involvement and visibility of the specific interaction structure can be treated as distinct design decisions. Whether the latter should be exposed should depend on its relevance to patient decision making, risk, and applicable professional or regulatory requirements. Similar questions warrant investigation in other high-stakes domains.

\subsection{Limitations and Future Work}

Some limitations qualify these conclusions. First, the experimental condition sizes were modestly unequal. Most cells contained approximately 27-29 participants, with one cell containing 22 (attributable to the \citet{kim2026implications} experiments). The Type II ANOVA framework accommodates unequal cell sizes, but the imbalance remains a feature of the design that should be considered when interpreting the estimates.

Second, although all other evaluation items were rated on 7-point scales, general trust was measured on a 1--100 scale to match the advising study~\cite{kim2026implications}, allowing the measure to be harmonized for the cross-domain analyses.

Third, our comparison includes only two expert domains: academic advising and medicine. Within medicine, however, the present study examines only a single, relatively structured DVT ultrasound workflow. The findings therefore should not be assumed to generalize to medical tasks that differ in procedural ambiguity, professional discretion, urgency, or degree of patient interaction. The absence of significant Domain $\times$ Performance interactions therefore does not establish that performance-related trust recalibration is universally domain-general. Nor do the nonsignificant interactions demonstrate statistical equivalence between the effects in the two domains. Instead, the present evidence is limited to showing that we detected no systematic domain moderation of the performance penalty across these two settings. Future work should extend the paradigm to domains such as finance, law, engineering, and other professional contexts in which expectations surrounding independent expertise, consultation, and technological assistance may differ.

Both studies relied on online participants evaluating simulated workflows. Participants recruited through Prolific were not making decisions as actual advisees or patients experiencing real consequences. Their evaluations may therefore differ from responses produced under greater emotional involvement, personal risk, or ongoing relationships with professionals. The medical setting in particular would benefit from replication in more realistic clinical simulations or field settings involving actual patients and healthcare professionals.

The simulation also contained fixed gender cues. Prior meta-analytic work found only a very small association between physician gender and patient satisfaction \cite{hall2011patients}, while experimental work on intelligent virtual assistants found little effect of voice gender alone on trust \cite{piercy2025gender}. The patient was not an evaluated actor and was shown only through the leg while participants imagined themselves in that role. Together, these considerations reduce concern that the fixed gender cues substantially shaped the observed effects.

There are also inferential limitations worth noting. While the full-model expertise Domain $\times$ Modality interaction survived multiple-testing corrections, the corresponding interaction in the AI-supported subset did not survive adjustment, and intended reuse proved sensitive to model specification. We therefore treat reuse effects as exploratory rather than confirmatory. Detailed explanations are provided in Appendix~\ref{app:limitations}.

Because AI-directed measures were structurally absent in the expert-only condition and represent distinct evaluative dimensions, trust toward the human versus the AI were evaluated via separate analyses rather than a formal Target $\times$ Performance interaction. Consequently, we interpret the result as an asymmetric empirical pattern rather than evidence that the human Performance effect was statistically larger than the AI effect. Further details are provided in Appendix~\ref{app:limitations}.

Future work should directly measure or manipulate the proposed mechanisms of AI-integration interpretation, including whether consultation is framed as routine professional verification or as assistance sought because of uncertainty, to test whether these perceptions mediate modality effects.
\section{Conclusion}

Across advising and medicine, observable expert failure broadly reduced human-facing evaluations, with no evidence that these penalties differed systematically by domain, while AI-directed trust remained comparatively insensitive. Expert failure therefore remained consequential even when AI support preserved a successful final outcome. AI integration structure was more selectively domain-sensitive: automatic oversight was associated with higher perceived expertise in medicine, and exploratory sensitivity analyses indicated a similar advantage for intended reuse after successful medical performance. In both patterns, expert-initiated AI did not differ detectably from expert-only, so the findings are better characterized as an automatic-oversight elevation than an invocation penalty. Together, the results distinguish comparatively portable performance-based recalibration from more context-sensitive responses to how AI enters expert workflows, underscoring that evaluation depends not only on what humans and AI do, but also on what their pattern of interaction communicates.

\section*{Acknowledgments}
\label{sec:ack}
Dr. Sreedharan's research is supported by NSF through grant 2303019. This material is based in part upon work supported by Other Transaction award 1AY2AX000062 from the U.S. Advanced Research Projects Agency for Health (ARPA-H) Platform Accelerating Rural Access to Distributed Integrated Medical Care (PARADIGM) program. The views and conclusions contained in this document are those of the authors and should not be interpreted as representing the official policies, either expressed or implied, of the U.S. Government.

\bibliographystyle{ACM-Reference-Format}
\bibliography{references}

\FloatBarrier

\appendix
\clearpage
\section{Supplementary Materials} \label{app:tables}
This appendix details the study simulation interface, supplementary statistical tables, and participant demographics.

\subsection{Simulation Interface} \label{app:user_study}

\begin{figure}[htbp]
    \centering
    \includegraphics[width=0.5\linewidth]{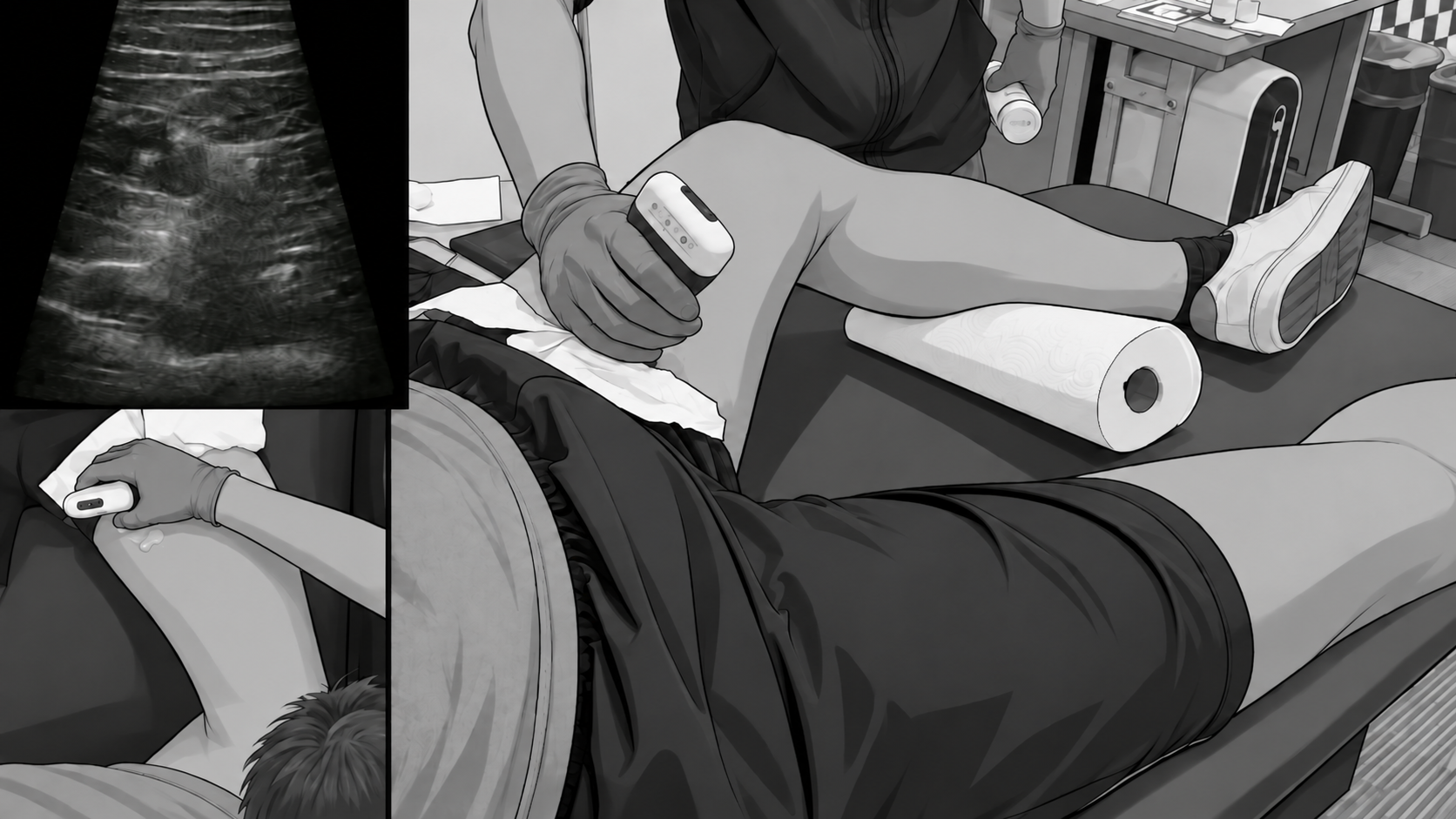}
    \caption{A representative frame from the simulated ultrasound scanning procedure observed by participants.}
    \Description{Representative frame from the simulated DVT ultrasound procedure. A technician operates an ultrasound probe on the patient's leg while viewing the scan on the ultrasound monitor.}
    \label{fig:scan}
\end{figure}

\begin{figure}[htbp]
    \centering
    \includegraphics[width=0.5\linewidth]{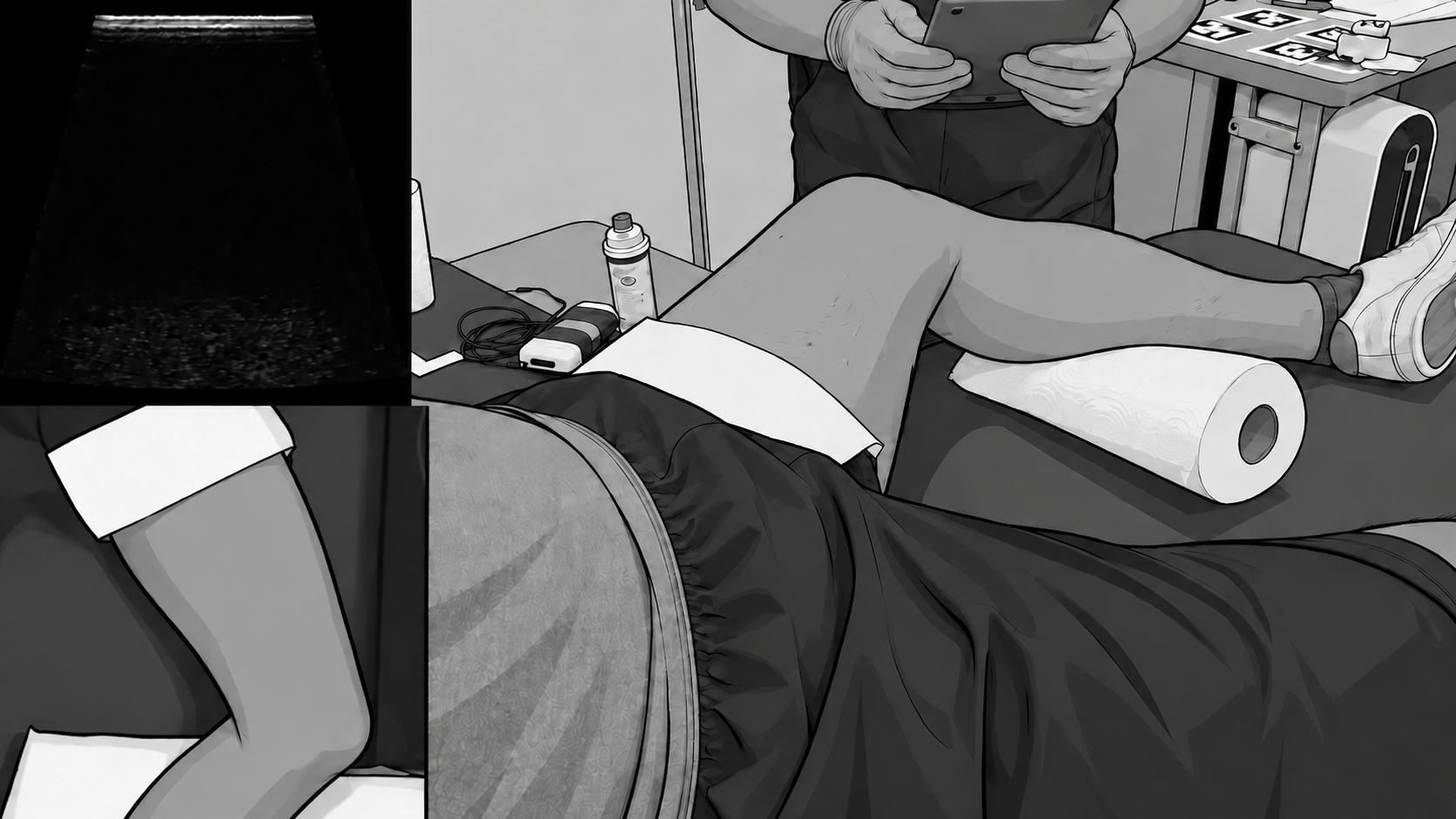}
    \caption{In the Expert-only condition, the technician checks scan accuracy against a reference book.}
    \Description{Expert-only condition showing the technician comparing the ultrasound scan with images in a reference book to check scan accuracy without AI assistance.}
    \label{fig:no_ai}
\end{figure}

Figure~\ref{fig:scan} shows a frame from the scanning procedure presented to participants during the simulation, where they were instructed to take the perspective of the patient. Figure~\ref{fig:no_ai} illustrates the checkpoint in the Expert-only condition, where the technician consults a reference book to verify that the scan has been performed correctly. Because the AI feedback in the AI-supported conditions was delivered via audio, no separate visual interface is depicted.

\subsection{Multiplicity Correction Families}
\label{app:multiplicity}

Multiplicity corrections were applied within predefined outcome families. Table~\ref{tab:multiplicity_families} reports the outcomes included in each family and the corresponding analysis scope. Corrections were applied separately within each family for each focal factorial effect using Benjamini-Hochberg false-discovery-rate (BH-FDR) correction, with Holm correction as a more conservative family-wise-error companion analysis.

\begin{table*}[t]
\centering
\small
\caption{Predefined outcome families used for across-outcome multiplicity correction.}
\Description{Table defining the three predefined outcome families used for across-outcome multiplicity correction. The shared human-facing family contains eight outcomes and was corrected separately within the full and AI-supported analyses. The AI-directed family contains three AI-focused trust outcomes and the team-directed family contains three human--AI team trust outcomes, both evaluated only in the AI-supported analysis.}
\label{tab:multiplicity_families}
\begin{tabular}{
    p{0.19\textwidth}
    p{0.20\textwidth}
    p{0.50\textwidth}
}
\toprule
\textbf{Outcome family} & \textbf{Analysis} & \textbf{Outcomes included} \\
\midrule

Shared human-facing (8)
&
Full $2 \times 3 \times 2$ and AI-supported $2 \times 2 \times 2$, corrected separately
&
METI expertise, integrity, and benevolence; expert perceived trustworthiness; expert affective comfort; expert information trust; general trust; and human-expert intended reuse. \\

\addlinespace

AI-directed (3)
&
AI-supported $2 \times 2 \times 2$ only
&
AI perceived trustworthiness; AI affective comfort; and AI information trust. \\

\addlinespace

Team-directed (3)
&
AI-supported $2 \times 2 \times 2$ only
&
Human-AI team perceived trustworthiness; human-AI team affective comfort; and human-AI team information trust. \\

\bottomrule
\end{tabular}

\vspace{0.5em}
\begin{minipage}{0.96\textwidth}
\footnotesize
\textit{Note.} The three expert-directed Riedl items are included within the shared human-facing family rather than treated as a separate outcome family. For each family, multiplicity correction was applied separately to the Performance, Domain $\times$ Performance, Domain $\times$ Modality, and Domain $\times$ Modality $\times$ Performance effects. BH-FDR correction was used as the primary across-outcome adjustment, with Holm correction applied to the same family as a more conservative family-wise-error companion analysis.
\end{minipage}
\end{table*}

\begin{table*}[t]
\centering
\small
\caption{Descriptive statistics (Mean ($SD$)) for medical-domain outcomes across Modality and Performance conditions. All outcomes were rated on 7-point scales (1-7), except general trust, which was rated on a 1-100 scale.}
\Description{Table reporting means and standard deviations for medical-domain outcomes across the six combinations of AI Modality and Performance. Outcomes are grouped into general, expert-directed, AI-directed, and expert--AI team-directed measures. Automatic AI generally showed the highest human-facing ratings, particularly in the success condition, including expertise, intended reuse, general trust, affective comfort, and information trust. AI- and team-directed outcomes were measured only in the two AI-supported conditions.}
\label{tab:descriptives}
\begin{tabular}{lcccccc}
\toprule
& \multicolumn{2}{c}{\textbf{Expert-only}} 
& \multicolumn{2}{c}{\textbf{Automatic AI}} 
& \multicolumn{2}{c}{\textbf{Expert-initiated AI}} \\
\cmidrule(lr){2-3} 
\cmidrule(lr){4-5} 
\cmidrule(lr){6-7}
\textbf{Outcome} 
& \textbf{Fail ($n=29$)} 
& \textbf{Succ ($n=27$)} 
& \textbf{Fail ($n=28$)} 
& \textbf{Succ ($n=28$)} 
& \textbf{Fail ($n=27$)} 
& \textbf{Succ ($n=27$)} \\
\midrule

\multicolumn{7}{l}{\textit{General Outcomes}} \\
Expertise score
& 4.43 (1.68) & 4.78 (1.50) & 5.26 (1.29) & 6.17 (1.12) & 4.57 (1.26) & 5.41 (1.38) \\
Integrity score
& 5.53 (1.30) & 5.84 (0.83) & 5.95 (0.94) & 6.27 (1.07) & 5.56 (0.92) & 6.19 (0.79) \\
Benevolence score
& 5.64 (1.30) & 5.93 (1.10) & 5.99 (1.09) & 6.12 (1.19) & 5.76 (0.98) & 5.95 (1.11) \\
Reuse (Expert)
& 4.07 (2.14) & 4.59 (1.89) & 4.86 (1.71) & 6.21 (1.40) & 4.26 (1.72) & 4.67 (2.15) \\
General trust
& 64.62 (29.41) & 69.33 (23.57) & 73.71 (22.49) & 86.46 (17.49) & 67.37 (20.14) & 69.78 (28.84) \\

\midrule
\multicolumn{7}{l}{\textit{Expert Outcomes}} \\
Trustworthiness (Expert)
& 5.00 (1.83) & 5.52 (1.40) & 5.54 (1.26) & 6.18 (1.06) & 5.19 (1.44) & 5.37 (1.74) \\
Affective comfort (Expert)
& 4.10 (2.06) & 4.56 (1.83) & 5.32 (1.49) & 5.93 (1.21) & 4.30 (1.81) & 5.07 (1.90) \\
Information trust (Expert)
& 4.69 (1.95) & 5.26 (1.72) & 5.82 (1.31) & 5.89 (1.37) & 4.81 (1.55) & 5.44 (1.48) \\

\midrule
\multicolumn{7}{l}{\textit{AI Assistant Outcomes}} \\
Trustworthiness (AI)
& -- & -- & 5.36 (1.66) & 5.14 (1.80) & 5.11 (1.62) & 4.67 (1.84) \\
Affective comfort (AI)
& -- & -- & 4.93 (1.90) & 4.86 (1.82) & 4.85 (1.77) & 4.44 (1.89) \\
Information trust (AI)
& -- & -- & 5.29 (1.72) & 5.00 (1.74) & 5.19 (1.44) & 4.67 (1.59) \\
Reuse (AI)
& -- & -- & 5.32 (1.68) & 5.32 (1.89) & 5.04 (1.79) & 4.74 (1.77) \\

\midrule
\multicolumn{7}{l}{\textit{Expert + AI Team Outcomes}} \\
Trustworthiness (Team)
& -- & -- & 5.57 (1.89) & 5.86 (1.60) & 5.30 (1.41) & 5.11 (1.72) \\
Affective comfort (Team)
& -- & -- & 5.39 (1.93) & 5.61 (1.71) & 4.85 (1.72) & 5.19 (1.73) \\
Information trust (Team)
& -- & -- & 5.61 (1.71) & 5.75 (1.46) & 5.11 (1.60) & 5.07 (1.80) \\
Reuse (Team)
& -- & -- & 5.21 (1.75) & 5.54 (1.88) & 4.44 (1.95) & 4.48 (1.97) \\

\bottomrule
\end{tabular}
\end{table*}

\begin{table*}[t]
\caption{Performance effects for the eight shared human-facing outcomes in the full
$2 \times 3 \times 2$ Type II ANOVAs ($N = 323$). Estimated marginal means average
equally across Domain and AI Modality. BH-FDR and Holm adjustments were applied
across the eight outcomes for the Performance effect.}
\label{tab:full_performance_effects}
\Description{Table reporting expert Performance effects for the eight shared human-facing outcomes in the combined advising and medical sample. For every outcome, the estimated marginal mean was lower following expert failure than following expert success. All eight Performance effects were statistically significant and remained significant after both BH-FDR and Holm correction. The table reports estimated marginal means, F statistics, p-values, partial eta-squared effect sizes, and adjusted p-values.}
\centering
\small
\begin{tabular}{lccccccc}
\toprule
& \multicolumn{2}{c}{Estimated marginal mean} & & & & & \\
\cmidrule(lr){2-3}
Outcome
& Failure
& Success
& $F(1,311)$
& $p$
& $\eta_p^2$
& $p_{\mathrm{BH}}$
& $p_{\mathrm{Holm}}$ \\
\midrule
Expertise
    & 5.05 & 5.81 & 29.74 & $<.001$ & .087 & $<.001$ & $<.001$ \\
Integrity
    & 5.68 & 6.10 & 15.98 & $<.001$ & .049 & $<.001$ & $<.001$ \\
Benevolence
    & 5.58 & 5.91 & 7.00 & .009 & .022 & .009 & .009 \\
Intended reuse
    & 4.63 & 5.64 & 27.86 & $<.001$ & .082 & $<.001$ & $<.001$ \\
General trust
    & 68.24 & 79.27 & 18.71 & $<.001$ & .057 & $<.001$ & $<.001$ \\
Trustworthiness (expert)
    & 5.29 & 5.90 & 14.87 & $<.001$ & .046 & $<.001$ & $<.001$ \\
Affective comfort (expert)
    & 4.76 & 5.63 & 22.50 & $<.001$ & .067 & $<.001$ & $<.001$ \\
Information trust (expert)
    & 5.21 & 5.87 & 16.50 & $<.001$ & .050 & $<.001$ & $<.001$ \\
\bottomrule
\end{tabular}
\end{table*}

\begin{table*}[t]
\caption{Entity-specific trust analyses in the AI-supported subset
($N = 219$). P = Performance; D$\times$P = Domain $\times$ Performance.}
\Description{Table reporting entity-specific Performance effects for trustworthiness, Affective comfort, and information trust in the AI-supported subset. Human-expert evaluations showed significant Performance effects for all three outcomes, with better evaluations following successful expert performance. AI-directed outcomes showed no significant Performance effects. Human-AI team outcomes also showed no significant Performance effects, and none of the Domain-by-Performance interactions for team-directed outcomes were significant.}
\label{tab:target_specific_performance}
\centering
\small
\begin{tabular}{lllccc}
\toprule
Target & Outcome & Effect & $F(1,211)$ & $p$ & $\eta_p^2$ \\
\midrule
Expert
    & Trustworthiness
    & P
    & 10.39 & .001 & .047 \\
Expert
    & Affective comfort
    & P
    & 15.88 & $<.001$ & .070 \\
Expert
    & Information trust
    & P
    & 10.79 & .001 & .049 \\
\addlinespace

AI
    & Trustworthiness
    & P
    & 0.02 & .884 & .000 \\
AI
    & Affective comfort
    & P
    & 0.30 & .582 & .001 \\
AI
    & Information trust
    & P
    & 0.01 & .935 & .000 \\
\addlinespace

Human-AI team
    & Trustworthiness
    & P
    & 2.06 & .153 & .010 \\
Human-AI team
    & Affective comfort
    & P
    & 3.52 & .062 & .016 \\
Human-AI team
    & Information trust
    & P
    & 3.15 & .078 & .015 \\
\addlinespace

Human-AI team
    & Trustworthiness
    & D$\times$P
    & 1.39 & .240 & .007 \\
Human-AI team
    & Affective comfort
    & D$\times$P
    & 0.43 & .515 & .002 \\
Human-AI team
    & Information trust
    & D$\times$P
    & 2.25 & .135 & .011 \\
\bottomrule
\end{tabular}
\end{table*}

\begin{table*}[t]
\centering
\scriptsize
\setlength{\tabcolsep}{3.5pt}
\caption{Exploratory fixed-cell modality contrasts for human-expert intended
reuse from the full ordinal-logit model. OR $>1$ favors the first-listed
modality for a higher reuse rating. $p_{\mathrm{Ord}}$, $p_{\mathrm{HC3}}$,
and $p_{\mathrm{MW}}$ are Holm-adjusted across the same 12 contrasts for,
respectively, the ordinal-Wald, HC3-linear, and Mann--Whitney analyses.}
\Description{Table reporting exploratory fixed-cell pairwise Modality contrasts for human-expert intended reuse across advising and medicine under failure and success conditions. Odds ratios from the ordinal-logit model are reported with 95\% confidence intervals, alongside Holm-adjusted p-values from ordinal-Wald, HC3-linear, and Mann-Whitney analyses. No contrasts were significant in advising or under medical failure. Under medical success, automatic AI produced significantly higher intended reuse than both expert-only and expert-initiated AI across all three analysis approaches.}
\label{tab:reuse_contrasts}

\begin{tabular}{lllccccc}
\toprule
Domain & Performance & Comparison
& OR & 95\% CI
& $p_{\mathrm{Ord}}$
& $p_{\mathrm{HC3}}$
& $p_{\mathrm{MW}}$ \\
\midrule

Advising & Failure
& Automatic AI vs. Expert-only
& 2.42 & [0.92, 6.37]
& .692 & .613 & .788 \\

Advising & Failure
& Expert-initiated AI vs. Expert-only
& 1.58 & [0.62, 4.02]
& 1.000 & 1.000 & 1.000 \\

Advising & Failure
& Automatic AI vs. Expert-initiated AI
& 1.54 & [0.60, 3.93]
& 1.000 & 1.000 & 1.000 \\

\addlinespace

Advising & Success
& Automatic AI vs. Expert-only
& 0.36 & [0.12, 1.08]
& .692 & 1.000 & .620 \\

Advising & Success
& Expert-initiated AI vs. Expert-only
& 0.51 & [0.17, 1.57]
& 1.000 & 1.000 & 1.000 \\

Advising & Success
& Automatic AI vs. Expert-initiated AI
& 0.70 & [0.27, 1.82]
& 1.000 & 1.000 & 1.000 \\

\addlinespace

Medicine & Failure
& Automatic AI vs. Expert-only
& 2.00 & [0.78, 5.11]
& 1.000 & 1.000 & 1.000 \\

Medicine & Failure
& Expert-initiated AI vs. Expert-only
& 1.19 & [0.47, 3.03]
& 1.000 & 1.000 & 1.000 \\

Medicine & Failure
& Automatic AI vs. Expert-initiated AI
& 1.68 & [0.68, 4.16]
& 1.000 & 1.000 & 1.000 \\

\addlinespace

Medicine & Success
& Automatic AI vs. Expert-only
& 7.77 & [2.84, 21.25]
& $<.001$ & .005 & .003 \\

Medicine & Success
& Expert-initiated AI vs. Expert-only
& 1.18 & [0.45, 3.09]
& 1.000 & 1.000 & 1.000 \\

Medicine & Success
& Automatic AI vs. Expert-initiated AI
& 6.59 & [2.36, 18.42]
& .004 & .024 & .011 \\

\bottomrule
\end{tabular}

\vspace{3pt}
\begin{minipage}{0.98\textwidth}
\footnotesize
\textit{Note.} The medical expert-initiated-AI success cell was highly
dispersed ($M=4.67$, $SD=2.15$, $n=27$): 7 of 27 responses were at 1--2
and 14 of 27 were at 6--7.
\end{minipage}
\end{table*}

\subsection{Participant Demographics} 
\label{app:demographics}

Table~\ref{tab:participant_demographics_wide} summarizes demographic characteristics and condition assignments for all participants in the medical simulation study ($N = 166$).

\begin{table*}[t]
\centering
\small
\caption{Participant Demographic Characteristics and Condition Allocations ($N = 166$)}
\Description{Table summarizing participant demographics and experimental condition allocations for the medical-domain sample of 166 participants. Demographic characteristics include gender, age group, education, and AI use frequency. The sample included 99 men, 63 women, and 4 non-binary participants, with the largest age group being 25--34 years and the most common education level being a bachelor's degree. Most participants reported using AI daily or weekly. Participants were distributed approximately evenly across the six Modality-by-Performance experimental conditions.}
\label{tab:participant_demographics_wide}
\begin{tabular}{lrr lrr lrr}
\toprule
\textbf{Category} & \textbf{\textit{n}} & \textbf{\%} & \textbf{Category} & \textbf{\textit{n}} & \textbf{\%} & \textbf{Condition} & \textbf{\textit{n}} & \textbf{\%} \\
\midrule
\multicolumn{3}{l}{\textit{Gender}} & \multicolumn{3}{l}{\textit{Education}} & \multicolumn{3}{l}{\textit{Experimental Conditions}} \\
Men        & 99 & 59.6\% & High school or equiv. & 30 & 18.1\% & Expert-only, Failure & 29 & 17.5\% \\
Women      & 63 & 38.0\% & Some college / univ.  & 26 & 15.7\% & Expert-only, Success & 27 & 16.3\% \\
Non-binary &  4 &  2.4\% & Associate degree      & 11 &  6.6\% & Automatic-AI, Failure     & 28 & 16.9\% \\
\cmidrule(r){1-3}
\multicolumn{3}{l}{\textit{Age Group}} & Bachelor's degree     & 71 & 42.8\% & Automatic-AI, Success     & 28 & 16.9\% \\
18--24     & 23 & 13.9\% & Master's degree       & 24 & 14.5\% & Expert-initiated-AI, Failure    & 27 & 16.3\% \\
25--34     & 57 & 34.3\% & Doctorate             &  4 &  2.4\% & Expert-initiated-AI, Success    & 27 & 16.3\% \\
\cmidrule(r){4-6}
35--44     & 47 & 28.3\% & \multicolumn{3}{l}{\textit{AI Use Frequency}} & & & \\
45--54     & 20 & 12.0\% & Daily                 & 89 & 53.6\% & & & \\
55--64     & 11 &  6.6\% & Weekly                & 49 & 29.5\% & & & \\
65+        &  8 &  4.8\% & Monthly               & 21 & 12.7\% & & & \\
           &    &        & Annual                &  7 &  4.2\% & & & \\
\bottomrule
\end{tabular}
\end{table*}

\subsection{Limitations}
\label{app:limitations}
Here, we discuss these inferential limitations in greater detail. Multiplicity strengthens the evidential distinction among the modality-related findings: the full-model expertise Domain $\times$ Modality interaction survived BH-FDR and Holm correction, whereas the corresponding AI-supported expertise interaction and the AI-supported reuse Domain $\times$ Modality effect did not. Intended reuse was also model-sensitive: the conventional three-way interaction missed the $0.05$ threshold, the full-model HC3 test crossed it while the AI-supported HC3 test did not, and cumulative-link analyses provided stronger evidence. Fixed-cell follow-ups nevertheless converged across methods. We therefore treat reuse as an exploratory sensitivity finding rather than a primary confirmatory result.

The cumulative-link models also assume common predictor effects across response thresholds, an assumption not independently tested here; agreement between logit and probit links does not establish that it holds. H2 was evaluated through separate human- and AI-target analyses rather than a formal Target $\times$ Performance interaction, so we interpret the result as an asymmetric empirical pattern rather than evidence that the human Performance effect was statistically larger than the AI effect. Nor did we directly measure the proposed mechanisms for modality interpretation, including uncertainty, dependence, responsibility, authority, consultation norms, or whether automatic intervention was perceived as institutional or background oversight. The qualitative observations therefore cannot establish mediation or causality. Finally, the shared ordering of expertise and successful-medical reuse does not establish a common psychological mechanism because their interaction structures differed.

\end{document}